\documentclass[pra,aps,twocolumn,amsfonts,showpacs]{revtex4}

\usepackage{graphicx}
\usepackage{epsfig}
\def\QED{\leavevmode\unskip\penalty9999 \hbox{}\nobreak\hfill
     \quad\hbox{\leavevmode  \hbox to.77778em{%
               \hfil\vrule   \vbox to.675em%
               {\hrule width.6em\vfil\hrule}\vrule\hfil}}
     \par\vskip3pt}
\def\qed{\leavevmode\unskip\penalty9999 \hbox{}\nobreak\hfill
     \quad\hbox{\leavevmode  \hbox to.77778em{%
               \hfil\vrule   \vbox to.675em%
               {\hrule width.6em\vfil\hrule}\vrule\hfil}}
     \par\vskip3pt}

\def\ibb #1{\leavevmode\hbox{\kern.3em\vrule
     height 1.5ex depth -.1ex width .4pt\kern-.3em\rm#1}}
\def\Cx {{\ibb C}}
\def\Rx {{\ibb R}}

\begin{document}

\title{Conditional entropies and their relation to entanglement criteria}

\author{Karl Gerd H. Vollbrecht}
\email{k.vollbrecht@tu-bs.de}
\author{Michael M. Wolf}
\email{mm.wolf@tu-bs.de} \affiliation{Institute for Mathematical
Physics, TU Braunschweig, Germany}

\date{\today}

\begin{abstract}
We discuss conditional R\'enyi and Tsallis entropies for bipartite
quantum systems of finite dimension. We investigate the relation
between the positivity of conditional entropies and entanglement
properties. It is in particular shown that any state having a
negative conditional entropy with respect to any value of the
entropic parameter is distillable since it violates the reduction
criterion. Moreover we show that the entanglement of Werner states
in odd dimensions can neither be detected by entropic criteria nor
by any other spectral criterion.
\end{abstract}


\maketitle

\section{Introduction}

Entanglement has always been a key issue in the ongoing debate
about the foundations and interpretation of quantum mechanics
since Einstein \cite{EPR} and Schr\"odinger \cite{Sch} expressed
their deep dissatisfaction about this astonishing part of quantum
theory. Whereas for the long period from 1935 to 1964, until Bell
\cite{Bell} published his famous work, discussions about
entanglement were purely meta-theoretical, nowadays quantum
information theory has established entanglement as a physical
resource and key ingredient for quantum computation and quantum
information processing. This led to a dramatic increase of general
structural knowledge about entanglement in the last few years, and
the resource point of view often led to results that are
reminiscent of those known from thermodynamics: {\it free
entanglement} is distinguished from {\it bound entanglement}
\cite{BE}, irreversibility can be observed in the process of
preparing and distilling entangled states \cite{irrev} and
entanglement itself is defined in a way that it must not increase
by means of local operations and classical communication (LOCC).
Moreover, there is recent effort in order to quantify quantum
correlations through heat engines \cite{heatengines}.

{\it Entropies} lay at the heart of both theories, thermodynamics
and entanglement theory. Concerning the latter it was shown that
few reasonable assumptions lead to a unique measure of
entanglement \cite{unique} for pure bipartite quantum states,
which is just the {\it von Neumann entropy} of the reduced state.
Hence, it is obvious that the two subsystems of a pure entangled
state exhibit more disorder as the system as a whole, such that
the respective conditional entropy is negative. This is a
remarkable property of entangled states, which is impossible for
classical systems (i.e. classical random variables).

The present paper is primarily devoted to settling the
relationship between the negativity of conditional R\'enyi and
Tsallis entropies and other entanglement properties. We will in
particular show how the property of having a positive conditional
entropy enters into the known implication chain of entanglement
resp. separability criteria.

In the second part we will then follow the result of Nielsen and
Kempe \cite{Nielsen} and give examples of entangled states having
the property that their entanglement can neither be detected by
entropic criteria nor by any other spectral criterion.
Sec.\ref{isosec} shows that this is indeed the case for symmetric
{\it Werner states} (in odd dimensions), which play a crucial role
in entanglement theory.

\section{Preliminaries on separability criteria}

To fix ideas we will start by recalling some of the basic notions
and previous results concerning separability resp. entanglement
criteria.

A bipartite quantum state described by its density matrix $\rho$
acting on a Hilbert space ${\cal H}={\cal H}^{(A)}\otimes{\cal
H}^{(B)}$ is said to be {\it separable}, unentangled or
classically correlated if it can be written as a convex
combination of tensor product states \cite{Werner89}
\begin{equation}\label{separable}
\rho = \sum_j p_j \rho_j^{(A)}\otimes\rho_j^{(B)},
\end{equation}
where the positive weights $p_j$ sum up to one and $\rho^{(A)}$
($\rho^{(B)}$) describes a state on ${\cal H}^{(A)}$ (${\cal
H}^{(B)}$). This means in particular that pure states are
separable if and only if they are product states. Moreover, all
entanglement properties of pure states, which can always be
written in their Schmidt form (cf. \cite{Schmidt}) as $
|\Psi\rangle = \sum_i \sqrt{\lambda_i} |i\rangle\otimes|i\rangle,
$  are completely determined by the eigenvalues $\{\lambda_i\}$ of
the reduced state $\rho_A={\rm tr}_B|\Psi\rangle\langle\Psi|$. The
unique measure of entanglement for pure states is then given by
the {\it von Neumann entropy} of the reduced state:
\begin{equation}\label{vNmeasure}
S_1(\rho_A)=-{\rm tr}\big(\rho_A\log\rho_A\big).
\end{equation}

For mixed quantum states however, the situation is much more
difficult and deciding whether a state is entangled or separable
is not yet feasible in general. Currently, the most efficient
necessary criterion for separability is the positivity of the
partial transpose (PPT), i.e., the condition that $\rho^{T_A}$ has
to be a positive semi-definite operator \cite{Peres}. The partial
transpose of the state is thereby defined in terms of its matrix
elements with respect to some basis by $\langle
kl|\rho^{T_A}|mn\rangle=\langle ml|\rho|kn\rangle$. For the
smallest non-trivial systems with $2\times2$ resp. $2\times3$
dimensional Hilbert spaces and a few other special cases
 the PPT-criterion also turned out to be
sufficient \cite{HHH}. In higher dimensional systems, however, so
called {\it bound entangled} states exist, which satisfy the
PPT-condition without being separable \cite{BE}.

Another well known condition is given by the {\it reduction
criterion} \cite{red,red2}
\begin{equation}\label{reduction}\rho_A\otimes{\bf
1}-\rho\geq 0,\quad\mbox{and}\quad {\bf 1}\otimes\rho_B-\rho\geq
0,
\end{equation}which is implied by the PPT-criterion but
nevertheless an important condition since its violation implies
the possibility of recovering entanglement by distillation (which
is yet unclear for PPT violating states).  For the case of two
qubits (and $2\times3$) the reduction criterion is also known to
be
 sufficient for separability \cite{red,red2}. Moreover, it was shown
in \cite{lowrank} that  Eq.(\ref{reduction}) implies that the rank
of the reduced state has to be smaller or equal than the rank of
$\rho$. The general line of implication is then:
\begin{equation}\label{ic}\begin{array}{c}
  \rho \mbox{ separable} \\\Downarrow\\ \rho^{T_A}\geq 0\ \\\Downarrow\\ \rho \mbox{ undistillable} \\\Downarrow\\ \rho_A\otimes{\bf
1}-\rho\geq 0\quad\wedge\quad {\bf 1}\otimes\rho_B-\rho\geq 0\\\Downarrow\\
\max\big[{\mbox{rank}(\rho_A),\mbox{rank}(\rho_B)\big]}\leq
\mbox{rank}(\rho)
\end{array}
   \end{equation}
The last condition we want to mention was recently derived by
Nielsen and Kempe \cite{Nielsen} and is based on majorization.
However, it is yet not known how the {\it majorization criterion}
enters into the above implication chain. Since it is closely
related to conditional entropies we will discuss it in more detail
in the following section.

\section{Conditional Entropies}\label{condent}

The idea to use entropic inequalities as separability resp.
entanglement criteria for mixed states goes back to the mid
nineties when Cerf and Adami \cite{Cerf} and the Horodecki family
\cite{HHHalpha} recognized that certain conditional R\'enyi
entropies are non-negative for separable states, and it was
recently resurrected by several groups
\cite{Abe1,Abe2,Lloyd,Vidiella,Rajagopal,Tsallisrecent} in the
form of conditional Tsallis entropies.

The quantum {\it R\'enyi entropy} depending on the entropic
parameter $\alpha\in\Rx$ is given by
\begin{equation}\label{Renyi}
S_\alpha (\rho) = \frac{\log {\rm tr}(\rho^\alpha)}{1-\alpha},
\end{equation}
where $S_0, S_1, S_\infty$ reduces to the logarithm of the rank,
the von Neumann entropy and the negative logarithm of the operator
norm respectively.  For the case of separable states it was shown
in \cite{Cerf,HHHalpha,lowrank} that the conditional entropy
\cite{secondred}:
\begin{equation}\label{Renyiineq}
S_\alpha(B|A;\rho):= S_\alpha(\rho)-S_\alpha(\rho_A)
\end{equation} is non-negative for
$\alpha=0,\infty$ and $\alpha\in[1,2]$.

In Ref. \cite{Abe1,Lloyd} essentially the same criterion was
expressed in terms of the {\it Tsallis entropy}
\begin{equation}\label{Tsallisent}
T_\alpha(\rho) = \frac{1-{\rm tr}(\rho^\alpha)}{\alpha-1},
\end{equation}which is non-negative, concave (convex) for $\alpha>0$
($\alpha<0$) and becomes the von Neumann entropy in the limit
$\alpha\rightarrow 1$. The {\it conditional Tsallis entropy}
defined in \cite{Abe1} reads
\begin{equation}\label{Tsalliscond}
T_\alpha(B|A;\rho):=\frac{{\rm tr}(\rho_A^\alpha)-{\rm
tr}(\rho^\alpha)}{(\alpha-1)\;{\rm tr}(\rho_A^\alpha)}\; .
\end{equation}
Concerning positivity, however, the two conditional entropies are
equivalent, i.e.
\begin{equation}\label{equi}
T_\alpha(B|A;\rho)\geq 0 \quad \Leftrightarrow \quad
S_\alpha(B|A;\rho)\geq 0 ,
\end{equation}
which is in turn equivalent to ${\rm tr}(\rho_A^\alpha)\geq{\rm
tr}(\rho^\alpha)$ for $\alpha> 1$, ${\rm
tr}(\rho_A^\alpha)\leq{\rm tr}(\rho^\alpha)$ for $0\leq\alpha< 1$,
and the positivity of the conditional von Neumann entropy for
$\alpha=1$.

Obviously, for pure states the conditional entropies are negative
if and only if the state is entangled.

\subsection{Monotonicity counterexample}\label{Monosec}

It was conjectured in \cite{Lloyd} that $T_\alpha(B|A;\rho)$ is
monotonically decreasing in $\alpha$, such that it would be
sufficient to calculate $T_\infty(B|A;\rho)$ in order to decide
positivity. However, monotonicity does not hold in general and can
most easily be ruled out by low rank examples like $$
\rho=\frac12\big(|\Phi_+\rangle\langle\Phi_+|+|01\rangle\langle
01|\big),\quad
|\Phi_+\rangle=\frac1{\sqrt{2}}\big(|00\rangle+|11\rangle\big), $$
for which the reduced state has eigenvalues $\frac14,\frac34$ and
therefore $T_0=T_\infty=0\neq T_2=\frac15$. We note that similar
 counterexamples can be found for the monotonicity of the conditional R\'enyi entropy as well. Fortunately, however, monotonicity is not
necessary for proving the positivity of the conditional
Tsallis/R\'enyi entropies for separable states for other values
than $\alpha=0,\infty$, $\alpha\in[1,2]$ \cite{wrongproof}.

\subsection{Majorization and convex functions}\label{Majo}

Majorization turned out to be a powerful tool in the discussion of
quantum state transformations by means of LOCC operations
(cf.\cite{majotrafo}) and it was recently proven to yield the
strongest separability criterion, which is based on the spectra of
a state and one of its reductions. It was proven in Ref.
\cite{Nielsen} that any separable state $\rho$ acting on
$\Cx^d\otimes\Cx^d$ is majorized by its reduced state $\rho_A$:

\begin{equation}\label{majo1}
\rho_A \succ \rho  \quad\mbox{ i.e. }\quad \forall  k\leq d :
\sum_{i=1}^k \lambda_i^{(A)} \geq \sum_{i=1}^k \lambda_i,
\end{equation}
where $\{\lambda_i\}$ and $\{\lambda_i^{(A)}\}$ are the
decreasingly ordered eigenvalues of $\rho$ respectively $\rho_A$.

It is a well known result in the theory of majorization that
$x\succ y$ iff ${\rm tr}\big(f(x)\big)\geq{\rm tr}\big(f(y)\big)$
for all convex functions $f:\Rx\rightarrow\Rx$ \cite{Bhatia}.
Since $f(x)=x^\alpha$ is convex for $\alpha\geq 1$, concave on
$\Rx^+$ for $0\leq\alpha\leq 1$ and the von Neumann entropy is
concave (needed for $\alpha=1$), this immediately implies:

\vspace{3pt}{\bf Theorem 1 }{\it Let $\rho$ be a bipartite quantum
state, which is majorized by its reduction $\rho_A\succ\rho$, then
for every $\alpha\geq 0$ the conditional Tsallis/R\'enyi entropies
of $\rho$ are non-negative, i.e.:
\begin{equation}\label{Eqtheo1}
  S_\alpha(B|A;\rho)\geq 0 \quad\mbox{and}\quad T_\alpha(B|A;\rho)\geq
  0.
\end{equation}
 }

The result of Nielsen and Kempe implies that this holds in
particular for any separable state.

 It is yet not known how the majorization criterion
 (\ref{majo1}) is related to other separability criteria like PPT,
 undistillability and the reduction criterion. However, we will
 show in the next subsection how the positivity of conditional
 entropies is related to these properties.

\subsection{Conditional Entropies and the Reduction
Criterion}\label{Redcrit}

Positivity of the conditional entropies for $\alpha=0$ reduces to
the rank criterion in the implication chain (\ref{ic}). The
following theorem will show, however, that all the other
properties stated in (\ref{ic}) in turn imply positivity of the
conditional entropies for every value of the entropic parameter
$\alpha$.

\vspace{3pt}{\bf Theorem 2 }{\it Let $\rho$ be a bipartite quantum
state satisfying the reduction criterion $\rho_A\otimes{\bf
1}\geq\rho$. Then for every $\alpha\geq 0$ the conditional
Tsallis/R\'enyi entropies are non-negative:
\begin{equation}\label{essence}
  S_\alpha(B|A;\rho)\geq 0 \quad\mbox{and}\quad T_\alpha(B|A;\rho)\geq
  0.
\end{equation}}\vspace{3pt}

We note that Thm.2 implies in particular, that states with
negative conditional entropies are distillable.

{\it Proof: } We will divide the proof into three steps depending
on the value of the entropic parameter.

 For {\bf$\alpha
> 1$} the proof is essentially based on the Golden-Thompson
inequality (cf.\cite{Petz}) stating that
\begin{equation}
{\rm tr}\big(e^A e^B\big) \geq {\rm tr}\big( e^{A+B}\big)
\end{equation}
for
 hermitian matrices $A,B$. Utilizing the
definition of the reduced state, i.e.,
\begin{equation}
\forall P\geq 0 : {\rm tr}\big(\rho(P\otimes{\bf 1} )\big)\equiv
{\rm tr}\big(\rho_A P \big)
\end{equation}
this leads to:
\begin{eqnarray}
{\rm tr}\big(\rho_A^\alpha\big) &=& {\rm
tr}\big[\rho(\rho_A^{\alpha-1}\otimes{\bf 1})\big]\nonumber\\ &=&
{\rm
tr}\Big[\exp\big(\ln\rho\big)\exp\big((\alpha-1)\ln(\rho_A\otimes{\bf
1} )\big)\Big]\nonumber\\ &\geq& {\rm
tr}\Big[\exp\big(\ln(\rho)+(\alpha-1)\ln(\rho_A \otimes{\bf 1}
)\big)\Big]\label{mono1}
\end{eqnarray}
At this point we need two monotonicity properties in order to
exploit the validity of the reduction criterion. First of all we
use the fact that the logarithm is operator monotone \cite{KL},
i.e.
\begin{equation}
A\geq B\Rightarrow\ln A\geq\ln B.
\end{equation}
Thus, for $\alpha > 1$ the reductions criterion $\rho_A \otimes
{\bf 1}\geq\rho$ implies
\begin{eqnarray}
\ln(\rho)+(\alpha-1)\ln(\rho_A \otimes{\bf 1})&\geq&
\ln(\rho)+(\alpha-1)\ln(\rho)\nonumber\\
&=&\alpha \ln (\rho).\label{g1}
\end{eqnarray}
In the second step we utilize the fact that the exponential
function is monotone under the trace.
 This can be seen by noting that for any $A$
hermitian, $P\geq 0$ and $B=(A+\epsilon P)$ with $\epsilon\geq 0$:
\begin{equation}\label{exp}
\frac{\partial}{\partial\epsilon} {\rm tr}\big(e^B\big)={\rm
tr}\big(e^B P\big)\geq 0.
\end{equation}
Hence ${\rm tr}\big(e^B\big)\geq{\rm tr}\big(e^A\big)$ is implied
by $B\geq A$. Together with Eq. (\ref{g1}) this leads to:
\begin{eqnarray}
{\rm tr}\big(\rho_A^\alpha\big)\geq (\ref{mono1})&\geq& {\rm
tr}\Big[\exp\big(\alpha\ln\rho\big)\Big]= {\rm
tr}\big(\rho^\alpha\big)\label{mono2}.
\end{eqnarray}

For $0\leq\alpha < 1$ the reduction criterion can immediately be
applied since $f(A)=A^r$ is an operator decreasing function for
$-1\leq r \leq 0,\; A\geq 0$ (cf.\cite{MO}) and thus
\begin{equation}\label{al1}
{\rm tr}\big(\rho_A^\alpha\big)={\rm
tr}\big[\rho(\rho_A^{\alpha-1}\otimes{\bf 1})\big]\leq {\rm
tr}\big(\rho^\alpha\big).
\end{equation}

For the case $\alpha=1$ we hav to look at the conditional von
Neumann entropy $S_1(\rho)- S_1(\rho_A)$, for which positivity is
directly implied by the reduction criterion and the operator
monotonicity of the logarithm:
\begin{eqnarray}
S_1(\rho_A) &=&-{\rm tr} \rho_A \log \rho_A \\
&=&-{\rm tr} \rho\log \rho_A \otimes {\bf 1} \\
&\leq&-{\rm tr} \rho\log \rho\\
&=& S_1(\rho) ,
 \end{eqnarray}
which completes the proof.

\QED

\subsection{Negative entropic parameters}\label{negalpha}
So far we have only discussed conditional entropies for
non-negative values of the entropic parameter $\alpha$. For these
cases we know that they can become negative for entangled states,
the simplest examples being pure entangled states. However, for
$\alpha<0$ (and states of full rank) the sign of the conditional
entropy contains no information:

\vspace{3pt}{\bf Theorem 3 }{\it Let $\rho$ be a bipartite quantum
state of full rank. Then for every $\alpha< 0$ the conditional
Tsallis/R\'enyi entropies are non-negative:
\begin{equation}\label{essencerank}
 \forall \alpha<0:\ S_\alpha(B|A;\rho)\geq 0 \quad\mbox{and}\quad T_\alpha(B|A;\rho)\geq
  0.
\end{equation}}\vspace{3pt}

{\it Proof: } Let  $\{|a\rangle \}$ be an eigenbasis of $\rho_A$.
Then:
 \begin{eqnarray}
 {\rm tr}\big(\rho^\alpha_A\big) &=& \sum_a \langle
 a|\rho_A|a\rangle^\alpha\\
 &=& \sum_a\Big[\sum_i\langle a\otimes i|\rho|a\otimes i\rangle\Big]^\alpha\label{f1}\\\
 &\leq& \sum_{a,i}\langle a\otimes i|\rho|a\otimes i\rangle^\alpha\leq
{\rm tr}\big(\rho^\alpha\big) ,\label{f2}
 \end{eqnarray}
 where Eq.(\ref{f1}-\ref{f2}) uses that $\big(\sum_i b_i\big)^\alpha\leq\sum_i b_i^\alpha$ holds for $b_i\geq 0,\; \alpha\leq0$, and the last inequality is implied by the convexity of negative
 powers on $\Rx^+$.

\QED

\section{Isospectral states}\label{isosec}

The fact that  positivity of conditional entropies is implied by
the reduction criterion (Thm.2) shows already that such an
entropic criterion cannot be sufficient for separability. In fact,
it was shown in Ref. \cite{Nielsen} that no spectral property is
capable of distinguishing any entangled state from separable ones.

We will in this section follow the idea of Ref. \cite{Nielsen} and
construct  particular examples of states, such that their
entanglement cannot be detected by any spectral criterion, since
there exist separable states having the same spectrum and the same
reductions.

{\it Werner states} \cite{Werner89} have always played an
important and paradigmatic role in quantum information theory.
Their characteristic property is that they commute with all
unitaries of the form $U\otimes U$ and they can be expressed as
\begin{equation}\label{Werner}
\rho(p)= (1-p)\frac{P_+}{r_+} + p \frac{P_-}{r_-} , \quad 0\leq p
\leq 1,
\end{equation}
where $P_+$ ($P_-$) is the projector onto the symmetric
(antisymmetric) subspace of $\Cx^d\otimes\Cx^d$ and $r_\pm = {\rm
tr}[P_\pm] = \frac{d^2\pm d}2$ are the respective dimensions.
Werner showed that these states are entangled iff $p>\frac12$
independent of the dimension $d$. The following shows however,
that none of these entangled states for odd dimension $d$ can be
detected by any separability criterion, which is based on the
spectrum of the state and its reductions.

\vspace{3pt}{\bf Theorem 4 }{\it Any entangled state in
$\Cx^d\otimes\Cx^d$ with maximal chaotic reductions and
eigenvalues having multiplicities which are multiples of $d$, has
a separable isospectral counterpart, which is locally
undistinguishable as it has the same reductions. }\vspace{3pt}

{\it Proof: }  Let us consider a special basis of maximally
entangled states in $\Cx^d\otimes\Cx^d$ \cite{W2}:
\begin{equation}\label{mebasis}
|\Psi_{jk}\rangle=\frac{1}{\sqrt{d}}\sum_{n=1}^d
\exp\Big(\frac{2\pi i}{d} j n \Big)|n, n\oplus k\rangle,
\end{equation}where $j,k=1,\ldots,d$ and $\oplus$ means addition
modulo $d$. Any equal weight combination of all states of the form
(\ref{mebasis}), which belong to the same value of $k$, is then a
projector onto a separable state since
\begin{eqnarray}\label{Psep}
P_k&=&\sum_{j=1}^d |\Psi_{jk}\rangle\langle\Psi_{jk}|\\
&=&\frac{1}{d}\hspace{-3pt}\sum_{j,n,m=1}^d\hspace{-4pt}
\exp\Big[\frac{2\pi i}{d} j(n-m) \Big] |n, n\oplus k\rangle\langle
m, m\oplus k|\nonumber\\ &=& \sum_{n=1}^d |n\rangle\langle
n|\otimes |n\oplus k\rangle\langle n\oplus k|\nonumber
\end{eqnarray}is an equal weight combination of product states.
Here we have used that $\frac1d \sum_{j=1}^d \exp\Big[\frac{2\pi
i}{d} j(n-m)\Big]=\delta_{n,m}$. Moreover, the reductions of the
respective states $P_k/d$ are maximally chaotic, i.e. $\rho_A={\bf
1}/d$, just as the reductions of any maximally entangled state.

If we now have a state with multiplicities being multiples of $d$
we can replace the projectors onto its eigenspaces with
sufficiently many projectors of the form $P_k$. The resulting
state will then be again a convex combination of product states,
i.e., separable, having the same spectrum and maximal chaotic
reductions.\QED

For the case of Werner states we note that the unitary invariance
of the state $\rho(p)$ in Eq. (\ref{Werner}) implies that its
reductions are $\rho_A={\bf 1}/d$. Moreover $\rho(p)$ has two
eigenvalues $(1-p)/r_+$ and $ p/r_-$ with multiplicities $r_+$,
$r_-$ which are indeed multiples of $d$ in odd dimensions.

Following Proposition 2 we can now construct a state
\begin{equation}\label{isowerner}
\rho'(p)=\frac{(1-p)}{r_+}\sum_{k=1}^{r_+/d} P_k + \frac{ p
}{r_-}\sum_{l=1}^{r_-/d} P_{l+r_+/d} ,
\end{equation} which has then both, the same spectrum and the same
reductions as $\rho(p)$. However, as convex combination of
separable states it is itself separable for any $0\leq p\leq 1$.

\section{Conclusion}

We discussed conditional R\'enyi and Tsallis entropies and the
relation between their positivity and other separability
properties. We showed in particular that states having a negative
conditional entropy are distillable since they violate the
reduction criterion.

Conditional entropies are a special instance of criteria using
just the spectra of a state and its reductions. Concerning the
detection of entanglement, it was shown in Ref.\cite{Nielsen} that
 majorization  is the strongest spectral criterion, which uses the
 spectra of a state and just one of its reductions. Its
 relation to other separability criteria is yet not known. The
 present result and numerical evidence may indicate that
 majorization is also implied by the reduction criterion. However,
 the proof presented in Sec.\ref{Redcrit} does not work for
 arbitrary convex functions and in fact majorization is not
 implied by the conditional entropy criteria.

Concerning separability the most  efficient  criterion is still
the PPT criterion, which is also a spectral criterion, however,
for the partially transposed state. One interesting question in
this context would therefore be: how can other (easy calculable)
invariants provide information about the separability of a state,
which is not yet encoded in the smallest eigenvalue of its partial
transpose?

\section*{Acknowledgement}
The authors would like to thank R.F. Werner for useful discussions
and M.A. Nielsen for bringing the relation to majorization to
their attention. Funding by the European Union project EQUIP
(contract IST-1999-11053) and financial support from the DFG
(Bonn) is gratefully acknowledged.


\begin{references}
\bibitem{secondred}Note that throughout this paper we omit equations corresponding to
the reduction with respect to the second subsystem $\rho_B={\rm
tr}_A(\rho)$. However, all the equations hold in a symmetric form
for both reductions and the presented ones should be read as
representatives for either of the two.
\bibitem{wrongproof}Abe and Rajagopal already tried to prove this proposition for separable states in
\cite{Abe1}. However, they failed since they wrongly supposed that
all the states in the decomposition in Eq.(\ref{separable})
mutually commute.
\end{references}
\end{document}